\documentclass[a4paper,11pt]{article}
\usepackage{jcappub} 
\usepackage{lineno}
\usepackage{natbib}
\usepackage[T1]{fontenc}
\usepackage{xcolor}
\usepackage{cleveref}
\usepackage{booktabs}
\usepackage{rotating}

\newcommand\aj{\ref@jnl{AJ}}
\newcommand\psj{\ref@jnl{PSJ}}
\newcommand\araa{\ref@jnl{ARA\&A}}
\newcommand\apj{\ref@jnl{ApJ}}
\newcommand\apjl{\ref@jnl{ApJL}}     
\newcommand\apjs{\ref@jnl{ApJS}}
\newcommand\ao{\ref@jnl{ApOpt}}
\newcommand\apss{\ref@jnl{Ap\&SS}}
\newcommand\aap{\ref@jnl{A\&A}}
\newcommand\aapr{\ref@jnl{A\&A~Rv}}
\newcommand\aaps{\ref@jnl{A\&AS}}
\newcommand\azh{\ref@jnl{AZh}}
\newcommand\baas{\ref@jnl{BAAS}}
\newcommand\icarus{\ref@jnl{Icarus}}
\newcommand\jaavso{\ref@jnl{JAAVSO}}  
\newcommand\jrasc{\ref@jnl{JRASC}}
\newcommand\memras{\ref@jnl{MmRAS}}
\newcommand\mnras{\ref@jnl{MNRAS}}
\newcommand\pra{\ref@jnl{PhRvA}}
\newcommand\prb{\ref@jnl{PhRvB}}
\newcommand\prc{\ref@jnl{PhRvC}}
\newcommand\prd{\ref@jnl{PhRvD}}
\newcommand\pre{\ref@jnl{PhRvE}}
\newcommand\prl{\ref@jnl{PhRvL}}
\newcommand\pasp{\ref@jnl{PASP}}
\newcommand\pasj{\ref@jnl{PASJ}}
\newcommand\qjras{\ref@jnl{QJRAS}}
\newcommand\skytel{\ref@jnl{S\&T}}
\newcommand\solphys{\ref@jnl{SoPh}}
\newcommand\sovast{\ref@jnl{Soviet~Ast.}}
\newcommand\ssr{\ref@jnl{SSRv}}
\newcommand\zap{\ref@jnl{ZA}}
\newcommand\nat{\ref@jnl{Nature}}
\newcommand\iaucirc{\ref@jnl{IAUC}}
\newcommand\aplett{\ref@jnl{Astrophys.~Lett.}}
\newcommand\apspr{\ref@jnl{Astrophys.~Space~Phys.~Res.}}
\newcommand\bain{\ref@jnl{BAN}}
\newcommand\fcp{\ref@jnl{FCPh}}
\newcommand\gca{\ref@jnl{GeoCoA}}
\newcommand\grl{\ref@jnl{Geophys.~Res.~Lett.}}
\newcommand\jcp{\ref@jnl{JChPh}}
\newcommand\jgr{\ref@jnl{J.~Geophys.~Res.}}
\newcommand\jqsrt{\ref@jnl{JQSRT}}
\newcommand\memsai{\ref@jnl{MmSAI}}
\newcommand\nphysa{\ref@jnl{NuPhA}}
\newcommand\physrep{\ref@jnl{PhR}}
\newcommand\physscr{\ref@jnl{PhyS}}
\newcommand\planss{\ref@jnl{Planet.~Space~Sci.}}
\newcommand\procspie{\ref@jnl{Proc.~SPIE}}

\newcommand\actaa{\ref@jnl{AcA}}
\newcommand\caa{\ref@jnl{ChA\&A}}
\newcommand\cjaa{\ref@jnl{ChJA\&A}}
\newcommand\jcap{\ref@jnl{JCAP}}
\newcommand\na{\ref@jnl{NewA}}
\newcommand\nar{\ref@jnl{NewAR}}
\newcommand\pasa{\ref@jnl{PASA}}
\newcommand\rmxaa{\ref@jnl{RMxAA}}

\title{A Multi‑Wavelength Survey of Transient Lensing Opportunities for Primordial Black Hole Searches}
\author{Stefano Profumo}
\affiliation{Santa Cruz Institute for Particle Physics and }
\affiliation{Department of Physics, University of California, Santa Cruz \\Santa Cruz, CA, 95064, USA}
\emailAdd{profumo@ucsc.edu}

\abstract{\noindent Gravitational lensing of short astrophysical transients provides a uniquely direct avenue for searching for primordial black holes (PBHs) across a vast range of masses. While past search efforts have focused on particular source classes—such as fast radio bursts (FRBs) and gamma-ray burst spikes—no systematic, multi-wavelength assessment has compared their relative potential for PBH discovery. We present here a broad assessment of transient lensing search opportunities, spanning more than twenty decades in photon frequency and over twelve orders of magnitude in PBH mass. For each class, we determine the accessible PBH mass window by accounting for wave–optics suppression and time–delay resolution limits, and we estimate potential  sensitivities to the PBH abundance using representative event rates, distances, and optical depths. Our survey includes low-frequency radio events (FRBs, pulsar giant pulses, planetary cyclotron bursts), optical/infrared signals, and high-energy phenomena (gamma-ray burst spikes, fast X-ray transients, TeV blazar flares). We synthesize these results in a unified mass–abundance diagram and comprehensive tables summarizing both physical reach and observational requirements. This work serves as a roadmap for optimizing future multi-wavelength lensing searches, guiding the design of instruments and strategies to explore the PBH dark matter hypothesis across its remaining viable parameter space.
}

\begin{document}

\maketitle
\flushbottom

\newpage
\section{Introduction}

Primordial black holes (PBHs) have long been recognized as a theoretically motivated and observationally elusive dark matter candidate, capable of spanning a wide range of masses from sub-asteroidal scales to many solar masses~\cite{Carr:1974nx,Hawking:1971ei,Carr:2020gox}.  
Although a diverse set of astrophysical and cosmological probes have excluded large fractions of this parameter space~\cite{Carr:2009jm,Green:2020jor}, substantial windows remain open.  
Identifying methods that can explore these gaps with minimal model dependence is therefore a central task in testing the PBH dark matter hypothesis, and for the field of dark matter searches more broadly intended.

One such method is gravitational lensing of short astrophysical transients.  
When a PBH lies close to the line of sight between an observer and a distant, compact source, it can produce detectable signatures: either two temporally separated images of the same burst, or interference fringes in the frequency spectrum caused by wave-optics effects~\cite{Paczynski:1986,Press:1973tv}.  
Because the basic physics depends only on the lens mass, geometry, and the properties of the light curve, transient lensing tests are among the cleanest probes of compact dark matter—requiring no assumptions about PBH formation history, accretion physics, or emission mechanisms.

While this concept is not new, prior work has concentrated on individual source classes: fast radio bursts (FRBs) as cosmological, high-rate probes~\cite{Munoz:2016tmg,Zhou:2022yxp}, giant pulses from pulsars for sub-microsecond timing~\cite{McKee:2018elc}, or gamma-ray burst spikes for high-energy, millisecond-scale events~\cite{Barnacka:2012bm}.  
Indeed, gravitational lensing of short-lived astrophysical transients has matured into a multi-faceted programme to probe PBHs across $\sim\!20$ orders of magnitude in mass.  
Because a transient's duration is usually far shorter than the light-travel time across a typical lens, even a single burst can reveal (or constrain) sub-arcsecond image separations through temporal or spectral structure that would be unobservable in quasi-steady sources such as quasars.

FRBs have emerged as a powerful new tool in the search for lensing effects that are sensitive to compact objects over a wide mass range~\citep{Munoz:2016tmg,Laha2018}. Their short duration and high-frequency radio signals make them ideal for detecting subtle time delays and interference patterns that would result from gravitational lensing by asteroid- to sub-solar-mass PBHs, particularly through wave-optics effects~\citep{Takahashi:2003ix}. Recent efforts, such as the CHIME/FRB baseband campaign, have pioneered the use of phase-coherent techniques to search for these effects, analyzing nanosecond-resolution voltage data from dozens of bursts~\citep{Leung2022,Kader2022}. While no definitive lensing signals have been observed so far, the results already place meaningful constraints on the PBH dark matter fraction at masses near \(10^{-3}\,M_\odot\), showing that even delicate phase information can survive propagation through the ionized intergalactic medium. Searches for lensing echoes at longer time delays---appearing as faint, secondary copies of the burst---have also been conducted using large FRB catalogues, placing weaker but steadily improving limits at higher PBH masses~\citep{Zhou2021}. Looking ahead, upcoming radio facilities such as DSA-2000, CHORD, and BURSTT, which will preserve the full electric field information of tens of thousands of FRBs, are expected to dramatically enhance the sensitivity of these searches~\citep{Connor2022}. 

Gamma-ray bursts have proven more challenging than initially anticipated.
A critical re-assessment showed that the projected size of the GRB photosphere typically exceeds the lens' Einstein radius, washing out fringes and invalidating earlier, superficially stringent asteroid-mass bounds \citep{Katz2018Femto}.  
The lightest PBH window therefore remains open unless future multi-satellite picolensing missions implement baselines of Earth–L2 or larger to recover sensitivity to $M\!\sim\!10^{-16}$–$10^{-12}\,M_\odot$ with parallax methods \citep{Fedderke2025}.
At the opposite extreme, several analyses of Fermi/GBM data have uncovered millisecond-to-tens-of-second echo pairs whose achromaticity and spectral similarity are consistent with $\sim\!10^{5}$–$10^{6}\,M_\odot$ intermediate-mass black-hole lenses \citep{Paynter2021,Kalantari2021,Lin2021}.  
So far each candidate stands alone, and population-level constraints remain highly uncertain because intrinsic GRB variability is a formidable contaminant, requiring uniform pipelines and end-to-end injection tests before these events can inform the PBH mass function.

Where FRBs and GRBs test coherence and temporal resolution, Type\,Ia supernovae leverage population statistics to constrain stellar-mass PBHs.  
In the geometric-optics regime, the probability distribution of magnifications is skewed by point-mass lenses, broadening the Hubble-diagram residuals.  
A joint analysis of the JLA and Union2.1 compilations constrains compact objects heavier than $\sim\!0.01\,M_\odot$ to contribute less than 35\% of dark matter at 95\% confidence level \citep{Zumalacarregui:2017qqd}.  
Because the source photosphere sets a hard lower mass bound, supernova microlensing complements rather than competes with FRB and GRB searches.  
Larger SN\,Ia samples expected from Rubin/LSST and Roman could soon push the stellar-mass PBH fraction below the tens-of-percent level.
Additionally, strongly lensed SNe\,Ia open a further avenue through differential microlensing of the multiple macro-images, which can isolate a compact lens population in the deflector galaxy.  
Simulation work suggests that even a handful of such systems could differentiate stellar microlenses from an $f_\mathrm{PBH}=1$ scenario in the 1–100\,$M_\odot$ range, although no statistical constraints have yet been published \citep{SunLiao2025}.

Gravitational-wave transients represent an emerging frontier for PBH searches, as chirping binary inspirals provide amplitude and phase information over approximately 2–3 decades in frequency.  
Wave-optics fringes can therefore imprint characteristic frequency-dependent modulations that diagnose compact lenses with $10\!\lesssim\!M/M_\odot\!\lesssim\!10^{5}$ \citep{JungShin2019}.  
First searches in the LIGO/Virgo O3 catalog saw no such distortions, limiting $f_\mathrm{PBH}$ to less than 0.5–0.8 around $10^{2}$–$10^{5}\,M_\odot$ under optimistic assumptions \citep{Basak2021}.
New lens stochastic diffraction formalisms promise order-of-magnitude sensitivity gains by folding in subtle amplitude fluctuations from an ensemble of faint images, potentially rivaling supernova bounds in the very-high-mass regime as Einstein Telescope and Cosmic Explorer come online \citep{Zumalacarregui2024LSD}.

The current state of the field reveals a clear division by mass scale. 
For sub-solar masses ($10^{-5}$–$10^{-1}\,M_\odot$), FRB coherent interference is the only demonstrated probe, already touching $f_\mathrm{PBH}\!\sim\!1$ at $M\!\sim\!10^{-3}\,M_\odot$ and poised for two-orders-of-magnitude improvement with high-frequency, voltage-preserving surveys.
At stellar masses ($10^{-1}$–$10^{2}\,M_\odot$), population statistics of SN\,Ia magnification currently set the tightest upper bound of $f_\mathrm{PBH}\!\lesssim\!0.35$, while FRB echo null searches are catching up and should become competitive with a ten-fold increase in well-sampled bursts.
For intermediate masses ($10^{3}$–$10^{6}\,M_\odot$), GRB millilensing candidates remain tantalising but not yet definitive, with gravitational-wave diffraction and future wide-field X-ray and gamma-ray missions providing the clearest path to robust limits.
Finally, at asteroid masses ($10^{-16}$–$10^{-12}\,M_\odot$), GRB femto- and picolensing remains theoretically viable only if source sizes are exceptionally small, with dedicated dual-spacecraft baselines offering a realistic strategy to resurrect sensitivity in this window.

Across all these observational classes, community consensus has shifted toward phase-preserving analyses, explicit modelling of finite-source and scattering decoherence, and end-to-end injection tests of detection efficiencies.  
With these lessons internalised, the next five years are likely to see FRBs deliver the first percent-level constraints over $10^{-4}$–$10^{2}\,M_\odot$, while supernovae and gravitational-wave methods cement the high-mass frontier.  
However, the rapid expansion of time-domain astronomy across the electromagnetic spectrum now presents a much broader landscape of potential targets beyond these established classes.
From nanosecond-resolved radio bursts to fast X-ray and gamma-ray transients, the diversity of known and hypothesized events spans more than twenty decades in photon frequency and offers sensitivity to PBHs over roughly fifteen orders of magnitude in mass.

The purpose of this work is to carry out the first systematic, multi-wavelength assessment of transient classes with potential for PBH lensing searches.  
Rather than setting new constraints, we focus on quantifying, comparing, and visualizing the discovery space each class can access.  
By incorporating both wave-optics suppression and time-delay resolution limits, and by combining them with realistic event rates and distances, we map each class onto a unified PBH mass–abundance diagram.  
This approach reveals not only where current searches are strongest, but also where underexplored transient populations could yield significant gains with targeted effort.

This paper is organized as follows.  
In Section~\ref{sec:theory}, we review the lensing formalism relevant for short-duration events.  
Section~\ref{sec:methods} describes our selection of transient classes, parameter assumptions, and methodology for computing both mass sensitivity ranges and forecasted abundance limits.  
In Section~\ref{sec:survey}, we present our multi-wavelength survey, detailing the properties and PBH reach of each class.  
Our results are synthesized in Section~\ref{sec:results} through comparative tables and a unified mass–abundance diagram.  
Section~\ref{sec:discussion} discusses the implications for optimizing future searches, and Section~\ref{sec:conclusions} summarizes our findings and outlines a roadmap for exploiting transient lensing across the electromagnetic spectrum. 
A final Appendix discusses in depth our assumptions for the calculation of the relevant optical depth for different transient classes.

\section{Theoretical Framework}
\label{sec:theory}

Gravitational lensing by a compact object of mass \(M\) produces multiple images of a background source if the impact parameter is sufficiently small compared to the Einstein radius~\cite{Paczynski:1986,Press:1973tv}.  
For short-duration transients, these images are typically unresolved spatially, but can be separated in \emph{time}, producing either distinct repeated pulses or interference fringes in the observed frequency spectrum.  
Both effects can be used to detect or constrain compact objects such as PBHs.

\subsection{Time Delay in Geometric Optics}

For a point-mass lens of mass \(M\), the gravitational time delay between the two lensed images is given by~\cite{Schneider:1992}
\begin{equation}
    \Delta t \simeq \frac{4GM}{c^3} \left(1 + z_L \right) \, F(y) ,
    \label{eq:tdelay}
\end{equation}
where \(G\) is Newton's constant, \(c\) is the speed of light, and \(z_L\) is the redshift of the lens.  
The function \(F(y)\) encodes the dependence on lensing geometry and is defined in terms of the normalized source–lens impact parameter \(y\), where
\begin{equation}
    y \equiv \frac{\beta}{\theta_E} ,
\end{equation}
with \(\beta\) the angular separation between the source and lens as seen by the observer, and \(\theta_E\) the angular Einstein radius:
\begin{equation}
    \theta_E = \sqrt{ \frac{4GM}{c^2} \, \frac{D_{LS}}{D_L D_S} } .
\end{equation}
Here, \(D_L\), \(D_S\), and \(D_{LS}\) are the angular diameter distances from observer to lens, observer to source, and lens to source, respectively.

The dimensionless time delay function \(F(y)\) takes the form
\begin{equation}
    F(y) = \frac{1}{2} y \sqrt{y^2 + 4} + \ln\left( \frac{\sqrt{y^2 + 4} + y}{\sqrt{y^2 + 4} - y} \right) .
\end{equation}
This function grows logarithmically for large \(y\) and approaches a finite value as \(y \to 0\), with \(F(1) \simeq 3.6\).

For typical near-alignment configurations with \(y \sim 1\), the delay scales approximately linearly with lens mass:
\begin{equation}
    \Delta t \sim 2 \times 10^{-5}~{\rm s} \left( \frac{M}{M_\odot} \right) (1+z_L) .
\end{equation}

The detectability of this delay in the \emph{geometric optics regime} requires that it exceed the intrinsic width of the transient:
\begin{equation}
    \Delta t \gtrsim \Delta t_{\rm burst},
    \label{eq:duration_condition}
\end{equation}
which sets a \emph{minimum lens mass} from the time-resolution requirement:
\begin{equation}
    M_{\rm min}^{\rm (duration)} \simeq \frac{c^3 \, \Delta t_{\rm burst}}{4 G (1+z_L)} .
    \label{eq:mmin_duration}
\end{equation}

\subsection{Wave--Optics and Finite-size-source Suppression}

For sufficiently low lens masses or high photon frequencies, the lensing phenomenon enters the \emph{wave--optics regime}, where diffraction effects smear out the interference pattern and suppress the lensing signal~\cite{Takahashi:2003ix,Nakamura:1998mt}.  
The relevant dimensionless parameter is
\begin{equation}
    w \equiv \omega \Delta t \approx \frac{4 G M \omega}{c^3} (1+z_L) ,
\end{equation}
where \(\omega = 2 \pi \nu\) is the angular frequency of the photons.  
The transition from wave– to geometric–optics occurs around \(w \sim 1\), giving a second minimum mass condition:
\begin{equation}
    M_{\rm min}^{\rm (wave)} \simeq \frac{c^3}{4 G \omega (1+z_L)} .
    \label{eq:mmin_wave}
\end{equation}

For a transient of given observing frequency \(\nu\) and duration \(\Delta t_{\rm burst}\), the overall minimum detectable PBH mass is
\begin{equation}
    M_{\rm min} = \max \left[ M_{\rm min}^{\rm (duration)}, \, M_{\rm min}^{\rm (wave)} \right] .
    \label{eq:mmin_total}
\end{equation}

In addition to wave effects, the finite angular size of the source can also suppress the lensing signal. This occurs when the source size projected onto the lens plane exceeds the Einstein radius of the lens, thereby washing out the interference fringes. The relevant condition can be quantified by the dimensionless source size parameter
\begin{equation}
    \rho \equiv \frac{R_S D_L}{r_E D_S} ,
\end{equation}
where \(R_S\) is the physical size of the source, \(D_L\) and \(D_S\) are the angular diameter distances to the lens and source, and \(r_E\) is the Einstein radius. Significant suppression occurs when \(\rho \gtrsim 1\), introducing an additional effective lower bound on the lens mass for a detectable signal. Finite-source effects are especially important for nearby lenses or extended emission regions, and must be accounted for in realistic lensing forecasts~\cite{Smyth:2019whb, Laha:2018zav}. Applications of this formalism to FRBs and other transients can be found in~\cite{Munoz:2016tmg,Zhou:2022yxp}. We discuss finite size source effects in the context of our multi-wavelength survey in sec.~\ref{sec:finite_size} below.

\subsection{Abundance Sensitivity}

If PBHs constitute a fraction \(f_{\rm PBH}\) of the dark matter, the optical depth for lensing in the geometric optics regime is~\cite{Niikura:2019kqi}
\begin{equation}
    \tau_{\rm lens}(M, f_{\rm PBH}) = f_{\rm PBH} \, \tau_{\rm lens}(M, f_{\rm PBH}=1) ,
\end{equation}
where \(\tau_{\rm lens}(f_{\rm PBH}=1)\) is the optical depth assuming PBHs make up all the dark matter.  
This depends on the cosmological geometry, source redshift distribution, and Einstein radius of the lens, but is only weakly dependent on \(M\) for fixed source distance in the point-lens limit.

If \(N_{\rm events}\) transients are observed, the absence of lensing detections sets an upper bound on \(f_{\rm PBH}\) at a chosen confidence level \(\alpha\) through~\cite{Zumalacarregui:2017qqd}
\begin{equation}
    f_{\rm PBH} \lesssim \frac{-\ln\alpha}{N_{\rm events} \, \tau_{\rm lens}(f_{\rm PBH}=1)} .
    \label{eq:fpbh_limit}
\end{equation}
Thus, the \(f_{\rm PBH}\) sensitivity improves linearly with event number, provided each sightline has comparable lensing probability and sufficient signal-to-noise.

\subsection{Summary of Key Dependencies}

The sensitivity of a transient class to PBHs is determined by:
\begin{itemize}
    \item \textbf{Minimum detectable mass:} Eq.~\eqref{eq:mmin_duration} (duration-limited) and Eq.~\eqref{eq:mmin_wave} (wave–optics-limited),
    \item \textbf{Abundance reach:} Eq.~\eqref{eq:fpbh_limit}, which scales inversely with the product of the event number and the optical depth,
\end{itemize}
In the next sections we apply this framework to a diverse set of transient classes, from low-frequency radio bursts to high-energy gamma-ray spikes, to map the discovery space for PBH lensing searches.


\begin{sidewaystable}[t]
\centering
\caption{Observational parameters and computed optical depths for each transient class, ordered by increasing observing frequency. 
Optical depths are calculated for $f_{\mathrm{PBH}}=1$ using the formulas in Appendix~\ref{app:optical_depth}, with class-specific characteristic distances and clumping factors $\delta$ where plausible.}
\label{tab:pbh_transients_parameters_updated}
\begin{tabular}{p{5cm} p{2cm} p{2cm} p{2cm} p{2cm} p{2cm} p{2cm} p{3cm}}
\hline
\textbf{Transient Class} & \textbf{Frequency [Hz]} & \textbf{Duration [s]} & \textbf{Events/year} & \textbf{Category} & \textbf{Distance} & \textbf{$\delta$} & \textbf{$\tau_{\mathrm{lens}}(f_{\mathrm{PBH}}=1)$} \\
\hline
Auroral kilometric radiation           & $10^{5}$                  & $10^{-6}$    & $10^{4}$ & local     & $3.3 \times 10^{-6}$ AU & 1.0 & $3.0 \times 10^{-37}$ \\
Exoplanet cyclotron bursts              & $2.5 \times 10^{7}$       & $10^{-6}$    & $10^{2}$ & local     & $10$ AU                 & 10.0 & $2.8 \times 10^{-23}$ \\
Solar decametric spikes                 & $10^{8}$                  & $10^{-5}$    & $10^{4}$ & local     & $1$ AU                  & 1.0 & $2.8 \times 10^{-26}$ \\
UCD/Brown dwarf cyclotron bursts        & $1.5 \times 10^{8}$       & $10^{-6}$    & $10^{2}$ & local     & $10^{5}$ AU ($\approx 0.5$ pc) & 5.0 & $1.4 \times 10^{-15}$ \\
Pulsar giant pulses (low-freq)          & $1.5 \times 10^{8}$       & $10^{-9}$    & $10^{3}$ & galactic & $2$ kpc                 & 5.0 & $1.6 \times 10^{-8}$ \\
FRB (current CHIME, worst case)         & $10^{9}$                  & $10^{-3}$    & $10^{4}$ & cosmo    & $z_s = 1.0$             & 1.0 & $10^{-3}$ \\
FRB (future SKA, best case)             & $10^{9}$                  & $10^{-3}$    & $10^{5}$ & cosmo    & $z_s = 1.0$             & 1.0 & $10^{-3}$ \\
GRB radio afterglow microstructure      & $10^{9}$                  & $10^{-6}$    & $10^{1}$ & cosmo    & $z_s = 1.0$             & 1.0 & $10^{-3}$ \\
Fast X-ray transients                   & $10^{18}$                 & $10^{-3}$    & $50$     & galactic & $8$ kpc                 & 1.0 & $2.6 \times 10^{-9}$ \\
Soft gamma repeater spikes              & $10^{20}$                 & $5.0 \times 10^{-4}$ & $5$ & galactic & $15$ kpc                & 5.0 & $9.2 \times 10^{-9}$ \\
GRB prompt spikes                       & $10^{20}$                 & $10^{-4}$    & $300$    & cosmo    & $z_s = 1.0$             & 1.0 & $10^{-3}$ \\
TeV blazar flares                       & $10^{26}$                 & $10^{0}$     & $20$     & cosmo    & $z_s = 0.5$             & 1.0 & $5.0 \times 10^{-4}$ \\
\hline
\label{tab:pbh_transients_parameters}
\end{tabular}
\end{sidewaystable}


\begin{sidewaystable}[t]
\centering
\caption{PBH mass sensitivity ranges and one-year $f_{\mathrm{PBH}}$ limits (95\% CL) for each transient class, ordered by increasing observing frequency. 
Values are computed using the optical depths in Table~\ref{tab:pbh_transients_parameters_updated}, with class-specific distances and clumping factors $\delta$ applied. 
$M_{\min}$ is the larger of the wave--optics and duration limits, $M_{\max}$ is set by the time--delay detectability ceiling.}
\label{tab:pbh_sensitivity_updated}
\begin{tabular}{p{5cm} p{3cm} p{3cm} p{3cm}}
\hline
\textbf{Transient Class} & \textbf{$M_{\min}$ [$M_\odot$]} & \textbf{$M_{\max}$ [$M_\odot$]} & \textbf{$f_{\mathrm{PBH}}$ limit (95\% CL)} \\
\hline
Auroral kilometric radiation           & $8.1 \times 10^{-2}$  & $10^{2}$ & $9.9 \times 10^{32}$ \\
Exoplanet cyclotron bursts              & $5.1 \times 10^{-2}$  & $10^{2}$ & $1.1 \times 10^{21}$ \\
Solar decametric spikes                 & $5.1 \times 10^{-1}$  & $10^{2}$ & $1.1 \times 10^{22}$ \\
UCD/Brown dwarf cyclotron bursts        & $5.1 \times 10^{-2}$  & $10^{2}$ & $2.1 \times 10^{13}$ \\
Pulsar giant pulses (low-freq)          & $5.4 \times 10^{-5}$  & $10^{2}$ & $1.9 \times 10^{5}$ \\
FRB (current CHIME, worst case)         & $8.1 \times 10^{-3}$  & $10^{2}$ & $3.0$ \\
FRB (future SKA, best case)             & $8.1 \times 10^{-3}$  & $10^{2}$ & $3.0 \times 10^{-1}$ \\
GRB radio afterglow microstructure      & $5.1 \times 10^{-5}$  & $10^{2}$ & $1.5 \times 10^{2}$ \\
Fast X-ray transients                   & $8.1 \times 10^{-3}$  & $10^{4}$ & $1.5 \times 10^{3}$ \\
Soft gamma repeater spikes              & $4.0 \times 10^{-3}$  & $10^{4}$ & $3.5 \times 10^{2}$ \\
GRB prompt spikes                       & $4.0 \times 10^{-5}$  & $10^{4}$ & $4.8$ \\
TeV blazar flares                       & $4.0 \times 10^{-7}$  & $10^{6}$ & $1.6 \times 10^{2}$ \\
\hline
\label{tab:pbh_transients_sensitivity}
\end{tabular}
\end{sidewaystable}

\section{Methodology}
\label{sec:methods}

Our goal is to compare the potential of different transient classes for detecting or constraining primordial black holes (PBHs) through gravitational lensing, across the widest possible mass range.  
To do so, we require for each class:
\begin{enumerate}
    \item The \emph{observational parameters} that determine the minimum detectable PBH mass (\S\ref{sec:methods:parameters}),
    \item The \emph{event rate}, \emph{typical source distance}, and \emph{clumping factor} that set the abundance sensitivity (\S\ref{sec:methods:rates}),
    \item The \emph{lensing optical depth} for the relevant geometry, computed from these distances and overdensities (\S\ref{sec:methods:tau}),
    \item The combination of these into a forecast for $f_{\mathrm{PBH}}$ limits and accessible mass range (\S\ref{sec:methods:calculation}).
\end{enumerate}

\subsection{Observational Parameters}
\label{sec:methods:parameters}

The mass reach for a given transient class is determined by two primary observables:
\begin{itemize}
    \item The \textbf{burst duration} $\Delta t_{\mathrm{burst}}$, setting the minimum mass via the time-delay resolution condition [Eq.~\eqref{eq:mmin_duration}],
    \item The \textbf{observing frequency} $\nu$, setting the minimum mass via the wave–optics condition [Eq.~\eqref{eq:mmin_wave}].
\end{itemize}
For each transient class, we adopt representative values for $\Delta t_{\mathrm{burst}}$ and $\nu$ from the literature or, where unavailable, from conservative physical estimates.  
In the FRB case, we use ms-scale widths from CHIME and ASKAP~\cite{CHIMEFRB:2021srp,Shannon:2018}.  
For pulsar giant pulses, we take ns–$\mu$s substructure from the Crab and PSR~B1937+21~\cite{Hankins:2003,Cordes:2004}.  
High-energy transient parameters are drawn from GRB prompt and afterglow studies~\cite{Fishman:1995,Meszaros:2006}, fast X-ray transients~\cite{Sguera:2006}, and TeV variability in blazars~\cite{Albert:2007}. 
Where possible, we use the shortest reliably observed substructure as $\Delta t_{\mathrm{burst}}$, rather than the overall event duration, since this is the relevant timescale for detecting lensing interference or repeats~\cite{Munoz:2016tmg,Zhou:2022yxp}.

\subsection{Event Rates, Source Distances, and Clumping Factors}
\label{sec:methods:rates}

The ability to constrain PBH abundance depends on the number of independent sightlines,
\begin{equation}
    N_{\mathrm{events}} = R_{\mathrm{sky}} \times T_{\mathrm{obs}} ,
\end{equation}
where $R_{\mathrm{sky}}$ is the annual all-sky event rate and $T_{\mathrm{obs}}$ is the observing duration (assumed to be 1~yr for forecasts).  
FRB rates are from CHIME/FRB~\cite{CHIMEFRB:2021srp} and SKA forecasts~\cite{Hashimoto:2020}, pulsar giant pulse rates from~\cite{Mickaliger:2012}, and GRB/X-ray/TeV rates from \emph{Swift}, \emph{Fermi}, and ground-based Cherenkov observatories~\cite{Lien:2016, Abeysekara:2018}.

The source distance $D_s$ strongly affects the optical depth. We adopt:
\begin{itemize}
    \item Cosmological transients: redshifts $z_s \simeq 0.5$–2, luminosity distances from a flat $\Lambda$CDM cosmology,
    \item Galactic sources: $D_s \sim$ few–tens of kpc,
    \item Local sources: $D_s$ from AU scales (solar) to $\lesssim$~pc (nearby brown dwarfs).
\end{itemize}

We also introduce a \textbf{clumping factor} $\delta \equiv \rho_{\mathrm{local}} / \bar{\rho}_{\mathrm{DM}}$ to account for possible PBH overdensities along specific sightlines (e.g., Galactic plane, stellar clusters, planetary systems).  
When appropriate, $\delta$ values of 5–10 are applied for Galactic and nearby sources.

\subsection{Optical Depth Calculation}
\label{sec:methods:tau}

For each class, the per-event lensing probability assuming $f_{\mathrm{PBH}}=1$ is computed as
\begin{equation}
\tau_{\mathrm{lens}} = \delta \int_{0}^{D_s} \frac{\rho_{\mathrm{DM}}(D)}{M_{\mathrm{PBH}}} \, \sigma_{\mathrm{lens}}(M_{\mathrm{PBH}}, D) \, dD ,
\end{equation}
where $\rho_{\mathrm{DM}}(D)$ is the dark matter density profile, $\sigma_{\mathrm{lens}}$ is the relevant cross-section in the geometric-optics regime (with wave–optics suppression applied near $M_{\min}$), and $\delta$ is the clumping factor.  
For cosmological sources, the integration is performed in comoving coordinates with $\rho_{\mathrm{DM}}(z) \propto (1+z)^3$; for Galactic and local sources, $D_s$ is small enough to treat $\rho_{\mathrm{DM}}$ as constant.

This procedure yields the $\tau_{\mathrm{lens}}$ values in Table~\ref{tab:pbh_transients_parameters_updated}, which now differ by many orders of magnitude between classes due to distance and $\delta$ effects — from $\simeq 10^{-3}$ for $z\sim 1$ FRBs and GRBs to $\lesssim 10^{-25}$ for Solar bursts.  
A detailed derivation of the optical depth expressions, including the geometric factors, cosmological integration limits, and the role of local overdensities, is provided in Appendix~\ref{app:tau_calc}.

\subsection{Calculation of Mass Ranges and Abundance Limits}
\label{sec:methods:calculation}

For each transient class, we compute:
\begin{enumerate}
    \item $M_{\min}$ from Eq.~\eqref{eq:mmin_total}, taking the larger of the duration and wave–optics limits,
    \item $M_{\max}$ from the scale where the lensing time delay exceeds both the burst duration and the maximum correlation window searched.
\end{enumerate}
For most ms–ns radio bursts, we adopt $M_{\max} \simeq 10^2\,M_\odot$; for longer-duration high-energy events, $M_{\max}$ can reach $10^{4}$–$10^{6}\,M_\odot$.

The 95\% CL abundance limits are forecast using
\begin{equation}
    f_{\mathrm{PBH,lim}} = \frac{-\ln(0.05)}{N_{\mathrm{events}} \, \tau_{\mathrm{lens}}(f_{\mathrm{PBH}}=1)} ,
\end{equation}
assuming zero detections and uniform sensitivity across events~\cite{Munoz:2016tmg,Zhou:2022yxp,Niikura:2019kqi}. Notice that while $f_{\rm PBH}$ is constrained by cosmology to be less than 1, for the purpose of comparing different transient classes we show results for $f_{\rm PBH}>1$ as well.



\section{Survey of Transient Classes}
\label{sec:survey}

We now summarize the astrophysical transient classes included in our multi‑wavelength survey.  
For each class, we describe the relevant phenomenology, typical observational parameters, representative event rates, and the corresponding PBH mass sensitivity and abundance reach derived from the methodology in Section~\ref{sec:methods}.  

\subsection{Low-Frequency Radio Transients}

\subsubsection*{Fast Radio Bursts (FRBs)}
FRBs are millisecond-duration, bright radio pulses of extragalactic origin, detected up to $z \gtrsim 1$~\cite{Lorimer:2007qn,CHIMEFRB:2021srp}.  
Their short durations, high fluence, and cosmological path lengths make them ideal for PBH lensing searches in the $8\times 10^{-3} \lesssim M_{\rm PBH}/M_\odot \lesssim 10^{2}$ range~\cite{Munoz:2016tmg,Zhou:2022yxp}.  
We adopt ms-scale burst durations, GHz frequencies, and event rates of $\sim 10^4$~yr$^{-1}$ for current CHIME and $\sim 10^5$~yr$^{-1}$ for future SKA-class surveys, giving $\tau_{\rm lens} \simeq 10^{-3}$ for $f_{\mathrm{PBH}}=1$.  
Our forecasts show FRBs set the strongest abundance limits in their mass window, reaching $f_{\mathrm{PBH}} \lesssim 0.3$ for SKA-scale rates and $\sim 3$ for current CHIME statistics.

\subsubsection*{Pulsar Giant Pulses}
Certain pulsars, such as the Crab and PSR~B1937+21, emit sporadic ``giant'' pulses with ns–$\mu$s substructure~\cite{Hankins:2003,Mickaliger:2012}.  
Observed at low frequencies ($\sim 150$~MHz), these can in principle probe PBHs down to $M_{\rm PBH} \sim 5\times 10^{-5} M_\odot$, slightly below the FRB limit due to their shorter durations.  
However, with Galactic distances ($D_s \sim 2$~kpc) and a clumping factor $\delta\simeq 5$, the optical depth remains small ($\tau_{\rm lens} \simeq 1.6\times 10^{-8}$), leading to $f_{\mathrm{PBH}}$ limits $\gtrsim 2\times 10^5$ for one year of observation, unless PBHs are strongly clustered in the relevant sightlines.

\subsubsection*{Planetary and Brown Dwarf Cyclotron Bursts}
Magnetized ultracool dwarfs (UCDs) and brown dwarfs emit coherent cyclotron maser bursts at tens to hundreds of MHz~\cite{Hallinan:2007}.  
Theoretical models predict analogous bursts from exoplanets~\cite{Zarka:2007}.  
We adopt $\mu$s-scale substructure for optimistic sensitivity.  
For nearby UCD/brown dwarfs ($D_s\sim 0.5$~pc, $\delta\simeq 5$), $M_{\min} \sim 0.05\,M_\odot$, but $\tau_{\rm lens} \simeq 1.4\times 10^{-15}$ makes one-year $f_{\mathrm{PBH}}$ limits extremely weak ($\sim 2\times 10^{13}$).  
For theoretical exoplanet cyclotron bursts at $\sim 10$~AU ($\delta\simeq 10$), $\tau_{\rm lens}$ is even smaller ($\sim 3\times 10^{-23}$), yielding $f_{\mathrm{PBH}}$ limits $\gtrsim 10^{21}$.

\subsubsection*{Solar Decametric Spikes}
The Sun produces intense, short-duration radio spikes at 10–300~MHz~\cite{Benz:1982}.  
While some reach tens of ms, we adopt optimistic $\Delta t_{\rm burst} \sim 10\ \mu$s for the mass-range calculation, giving $M_{\min} \sim 0.5\,M_\odot$.  
Their Solar System location ($\tau_{\rm lens} \simeq 3\times 10^{-26}$) makes them unsuitable for meaningful abundance constraints, with implied one-year limits $f_{\mathrm{PBH}} \gtrsim 10^{22}$.

\subsection{High-Energy and Other Transients}

\subsubsection*{Gamma-Ray Burst (GRB) Prompt Spikes}
GRBs exhibit rapid substructure in their prompt keV–MeV emission, with spikes as short as $\lesssim 100\ \mu$s~\cite{Fishman:1995,Meszaros:2006}.  
At $\nu \sim 10^{20}$~Hz, wave–optics suppression sets $M_{\min} \simeq 4\times 10^{-5} M_\odot$.  
Cosmological distances yield $\tau_{\rm lens} \simeq 10^{-3}$, but low annual event rates ($\sim 300$ yr$^{-1}$) limit $f_{\mathrm{PBH}}$ sensitivity to $\gtrsim 5$.

\subsubsection*{GRB Radio Afterglow Microstructure}
Early-time radio afterglows of GRBs may show ms–$\mu$s variability at $\nu \sim 1$~GHz~\cite{Frail:1997}.  
If confirmed and monitored, these could probe PBHs in the FRB-like mass range with similar cosmological path lengths, but the rarity of events ($\sim 10$ yr$^{-1}$) and $\tau_{\rm lens} \simeq 10^{-3}$ makes one-year limits noncompetitive ($f_{\mathrm{PBH}}\gtrsim 1.5\times 10^2$).

\subsubsection*{Soft Gamma Repeater (SGR) Spikes}
Magnetar giant flares produce ms- to sub-ms-scale spikes in the hard X-ray band~\cite{Hurley:2005}.  
For Galactic sources at $D_s\sim 15$~kpc and $\delta\simeq 5$, we find $\tau_{\rm lens} \simeq 9.2\times 10^{-9}$.  
Given the rarity ($\lesssim 5$ per year), this yields poor abundance sensitivity ($f_{\mathrm{PBH}} \gtrsim 3.5\times 10^2$) despite $M_{\min} \simeq 4\times 10^{-3} M_\odot$.

\subsubsection*{Fast X-ray Transients (FXTs)}
FXTs are sub-second flares in the keV band detected by missions such as INTEGRAL and \emph{Swift}~\cite{Sguera:2006}.  
We adopt $\Delta t_{\rm burst} \sim 1$~ms and $\nu \sim 10^{18}$~Hz, giving $M_{\min} \simeq 8\times 10^{-3} M_\odot$ and $M_{\max} \simeq 10^{4} M_\odot$.  
Galactic distances ($D_s\sim 8$~kpc) yield $\tau_{\rm lens} \simeq 2.6\times 10^{-9}$, and with $\sim 50$ events per year, one-year limits are weak ($f_{\mathrm{PBH}}\gtrsim 1.5\times 10^3$).

\subsubsection*{TeV Blazar Flares}
Some blazars exhibit fast TeV variability on timescales of minutes~\cite{Albert:2007}.  
At $\nu \sim 10^{26}$~Hz, only very massive PBHs ($\gtrsim 4\times 10^{-7} M_\odot$) produce detectable delays, with $M_{\max} \simeq 10^6 M_\odot$.  
For $z_s\simeq 0.5$, $\tau_{\rm lens} \simeq 5\times 10^{-4}$, and $\sim 20$ events per year, the one-year limit is $f_{\mathrm{PBH}}\gtrsim 1.6\times 10^2$.
\subsection{Finite Source Size Effects}
\label{sec:finite_size}

In the standard point-source approximation, gravitational lensing is computed under the assumption that the source emission region is spatially unresolved with respect to the angular scale set by the Einstein radius.  
This approximation breaks down when the projected size of the source on the lens plane becomes comparable to or exceeds the Einstein radius. In this regime, wave interference patterns are smeared out, geometric time delays are partially washed out, and the magnification signal is suppressed.  
Here we assess the relevance of finite-source-size effects for each transient class considered in this work, using approximate emission-region scales drawn from the literature or inferred from variability timescales.

\paragraph{Fast Radio Bursts (FRBs):}  
FRBs are among the most compact astrophysical transients, with brightness temperatures exceeding \(10^{35}\,\mathrm{K}\) and inferred emission regions of order tens to hundreds of kilometers~\cite{Cordes:2019}.  
At cosmological distances, such sizes are orders of magnitude smaller than the projected Einstein radius for PBHs in the mass range \(M_{\mathrm{PBH}} \gtrsim 10^{-6}\,M_\odot\).  
Finite source size effects are thus negligible for FRB lensing in the parameter space considered here.

\paragraph{Pulsar Giant Pulses:}  
Giant pulses from millisecond pulsars are similarly compact, with nanosecond-scale structures implying emission regions as small as \(\sim 1\,\mathrm{km}\).  
Given their Galactic distances (\(\sim\) kpc) and the same lensing scale, these sources also fall well within the point-source regime.

\paragraph{GRB Radio Afterglow Microstructure:}  
Radio microstructure in GRB afterglows likely arises from small-scale inhomogeneities in the external shock or the emission region.  
While the entire afterglow image can be large (\(\sim 10^{16}\,\mathrm{cm}\) after days), microsecond features imply localized structures of \(\lesssim 10^{10}\,\mathrm{cm}\), or \(\lesssim 0.1\,\mathrm{AU}\).  
At redshifts \(z\sim1\), this is still small compared to the Einstein radius for \(M_{\mathrm{PBH}} \gtrsim 10^{-2}\,M_\odot\), making lensing signatures potentially detectable provided the time resolution is sufficient.

\paragraph{Gamma-Ray Bursts (GRBs) and FXTs:}  
Prompt GRB spikes and fast X-ray transients involve relativistically beamed emission regions with variability timescales down to milliseconds or shorter.  
The associated emission regions are again \(\lesssim 10^8\)–\(10^9\,\mathrm{cm}\), i.e., much smaller than the Einstein diameter for lens masses \(M \gtrsim 10^{-3}\,M_\odot\) and source–lens geometries relevant to our analysis.  
Finite size effects are thus subdominant.

\paragraph{TeV Blazar Flares:}  
Blazar flare regions, particularly at TeV energies, are often modeled as compact emission zones near the base of the jet, with sizes constrained by day-scale variability to be \(\lesssim 10^{15}\,\mathrm{cm}\).  
This corresponds to \(\sim 10^{-3}\,\mathrm{pc}\), which is still much smaller than the projected Einstein radius for lens masses \(M_{\mathrm{PBH}} \gtrsim 10^{-4}\,M_\odot\), even for sources at \(z \sim 0.1\)–1.  
Thus, finite source size is unlikely to degrade lensing signatures in this context.

\paragraph{Solar Radio Spikes:}  
These coherent emissions occur in the solar corona, with emitting regions often extending over \(\gtrsim 10^8\)–\(10^9\,\mathrm{cm}\), and located at distances of \(\sim 1\,\mathrm{AU}\).  
Given the extremely small source–lens distance and modest lens masses probed, the Einstein radius projected onto the source plane is often \(\sim 10^8\,\mathrm{cm}\) or smaller.  
Thus, finite source size effects can substantially suppress lensing signatures in this case unless the PBHs are extremely nearby or the emission region is unusually compact.

\paragraph{Planetary and Ultracool Dwarf (UCD) Cyclotron Bursts:}  
These auroral or cyclotron emissions can span large magnetospheric scales—e.g., \(\sim R_J\) for Jupiter.  
Given their proximity (\(<10\,\mathrm{pc}\)) and the relatively small PBH masses (\(\lesssim M_\odot\)) relevant for such short-duration bursts, the Einstein radius can be \(\lesssim 10^9\,\mathrm{cm}\).  
Since the emission regions are likely comparable or larger, finite size effects may smear out lensing signatures, especially for higher mass PBHs or slower bursts.

\paragraph{Auroral Kilometric Radiation (AKR):}  
AKR emission originates from regions several Earth radii in extent (\(\sim 10^9\,\mathrm{cm}\)), and the source–observer distance is only about \(1\,\mathrm{AU}\).  
For lens masses up to \(\sim 1\,M_\odot\), the Einstein radius is comparable or smaller than the source size, leading to strong finite source size suppression.  
Thus, this class is largely insensitive to lensing in the standard picture unless the emission region is unexpectedly compact or PBHs are extremely close and clumped.\\

In summary, finite source size effects are negligible for extragalactic sources with microsecond–millisecond variability (FRBs, GRBs, blazars) and for Galactic giant pulses.  
They become relevant or dominant for local low-frequency bursts (solar, planetary, and auroral), where large emission regions and small Einstein radii combine to suppress lensing signatures.  
For completeness, detailed modeling of source geometries and emission coherence scales could be incorporated in future refinements of the survey sensitivity.

\begin{figure}[t]
    \centering
    \includegraphics[width=0.85\textwidth]{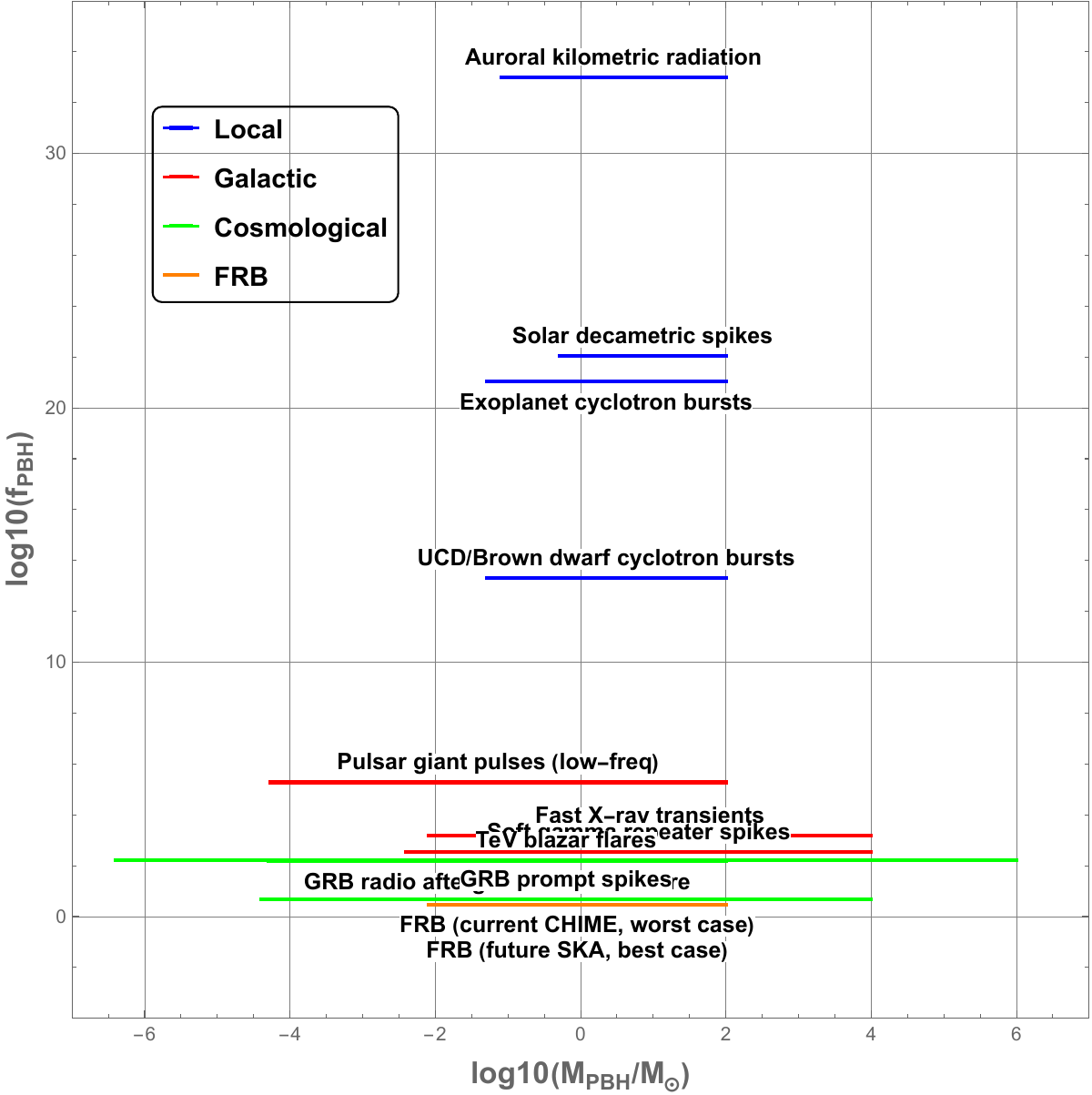}
    \caption{
        Unified PBH mass–abundance diagram for the transient classes in our survey.
        Symbols mark the minimum detectable $M_{\mathrm{PBH}}$ and corresponding $f_{\mathrm{PBH}}$ forecast for one year of observations.
        The horizontal bars indicate the approximate sensitivity range in PBH mass for each class.
        Orange: FRBs; blue: local low‑frequency transients; red: Galactic transients; green: high‑energy transients.
    }
    \label{fig:mass_abundance}
\end{figure}

\begin{figure}[t]
    \centering
    \includegraphics[width=0.85\textwidth]{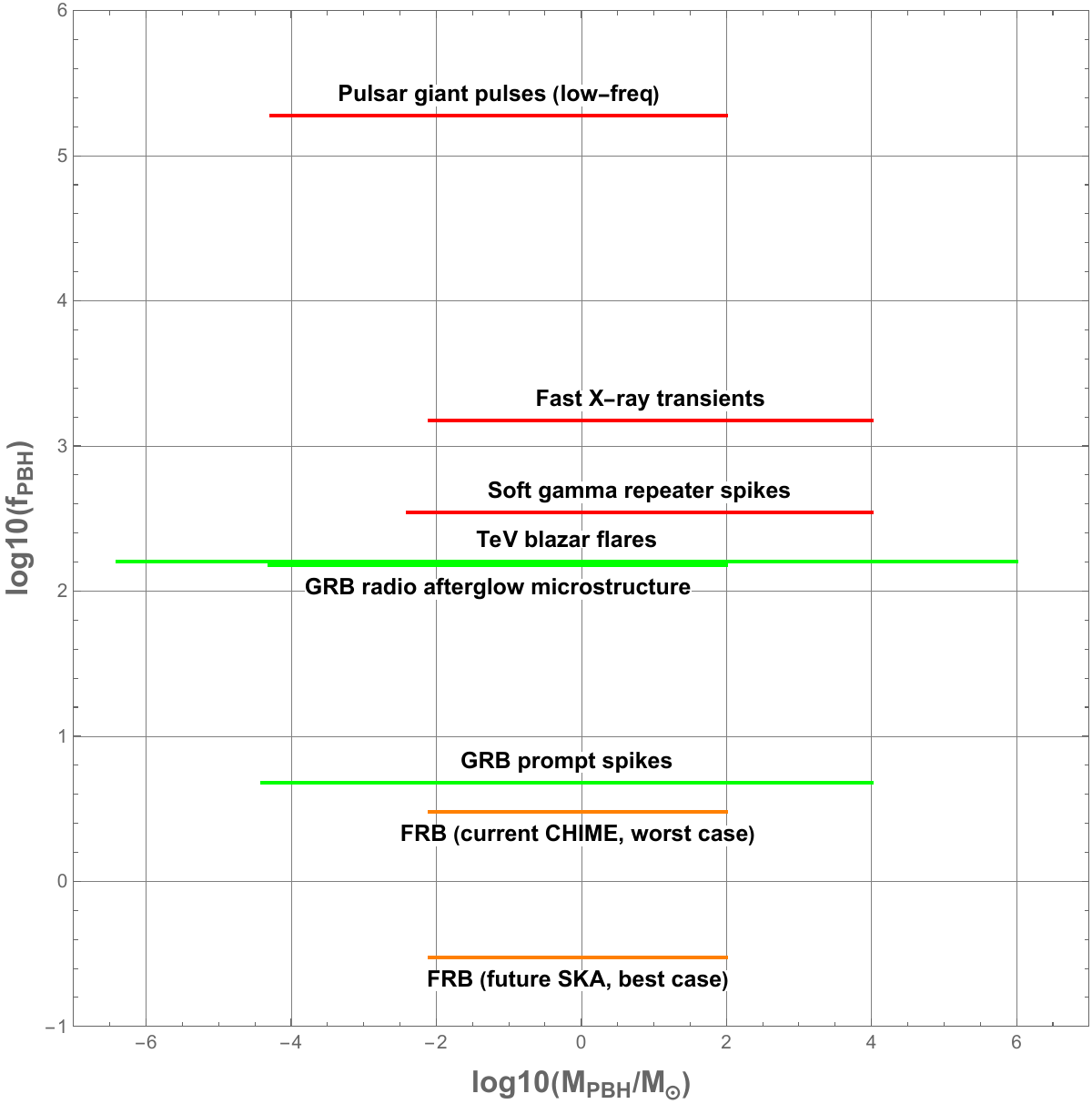}
    \caption{
        As in Fig.~\ref{fig:mass_abundance}, but restricted to $f_{\mathrm{PBH}}\lesssim 10^3$ to emphasize differences among the more sensitive classes.
    }
    \label{fig:mass_abundance2}
\end{figure}

\section{Results}
\label{sec:results}

In this section we present the quantitative outputs of our multi‑wavelength transient lensing survey.  
The methodology described in Section~\ref{sec:methods} is applied to the  observational parameters in Table~\ref{tab:pbh_transients_parameters_updated}, yielding revised mass sensitivity ranges, one‑year $f_{\mathrm{PBH}}$ limits, and qualitative sensitivity labels in Table~\ref{tab:pbh_sensitivity_updated}.  
We also synthesize these outputs in a unified mass–abundance diagram (Figures~\ref{fig:mass_abundance}-\ref{fig:mass_abundance2}) showing the coverage and relative sensitivity of each transient class.


Table~\ref{tab:pbh_sensitivity_updated} lists, for each transient class:
\begin{itemize}
    \item The minimum detectable PBH mass $M_{\min}$ from the larger of the wave–optics and duration limits,
    \item The adopted $M_{\max}$ defining the upper end of the sensitivity range,
    \item The forecast 95\% CL $f_{\mathrm{PBH}}$ limit for one year of observations with zero detections.
\end{itemize}
%
%
Our findings systematically evaluate the capacity of diverse astrophysical transients to constrain the fraction of dark matter composed of primordial black holes (PBHs), using a unified formalism that incorporates wave–optics and geometric delay constraints, realistic event rates, lensing optical depths, and finite source size effects. The results span over fourteen orders of magnitude in PBH mass and more than twenty decades in photon frequency, as visualized in the unified mass–abundance diagrams (Figures~\ref{fig:mass_abundance} and~\ref{fig:mass_abundance2}) and quantified in Table~\ref{tab:pbh_sensitivity_updated}.

Fast radio bursts (FRBs) emerge as the most powerful single class of probes, combining millisecond durations, cosmological distances, and large expected all-sky rates. In the best-case scenario involving future SKA-class instruments, FRBs can probe PBH masses in the range $8 \times 10^{-3}$ to $10^2\,M_\odot$ and reach projected one-year constraints of $f_{\mathrm{PBH}} \lesssim 0.3$. Current-generation FRB surveys such as CHIME are already providing partial coverage of this regime, although with correspondingly weaker sensitivity due to lower event rates.

Pulsar giant pulses, especially at low radio frequencies, access even lighter PBH masses than FRBs due to more favorable wave–optics scaling. However, their utility is limited by very small lensing optical depths and modest event rates. The best-case one-year constraint from this class remains at $f_{\mathrm{PBH}} \gtrsim 10^5$, making it far less competitive unless new high-rate sources are identified or substantially more sensitive campaigns are developed.

GRB radio afterglow microstructure provides sensitivity to similar PBH mass ranges as pulsar pulses, reaching $M_{\min} \sim 5 \times 10^{-5}\,M_\odot$, though with somewhat improved constraints on $f_{\mathrm{PBH}}$ due to more favorable event parameters. Nevertheless, limitations in time resolution and source rate still keep the achievable limits above $f_{\mathrm{PBH}} \sim 10^2$.

High-energy transients such as GRB prompt spikes, fast X-ray transients, soft gamma repeater bursts, and TeV blazar flares extend the accessible mass window up to $\sim 10^6\,M_\odot$. These sources, due to their cosmological distances and short durations, offer $\mathcal{O}(1)$ optical depths, but typically have lower detection rates. As a result, their one-year constraints on $f_{\mathrm{PBH}}$ generally range from a few hundred to unity, with the most favorable cases being GRB prompt spikes and afterglows. Their role is therefore to bridge the mass gap between radio and optical microlensing surveys and gravitational wave merger constraints.

Finally, local low-frequency bursts—including solar radio spikes, planetary cyclotron emission, and brown dwarf bursts—nominally probe intermediate PBH masses but are overwhelmingly limited by the extremely low optical depth in the local volume. Even under optimistic assumptions for PBH clumping, these events yield $f_{\mathrm{PBH}}$ limits in excess of $10^{20}$ or even $10^{30}$, making them ineffective for meaningful dark matter constraints unless substructure in the local halo is much more pronounced than currently believed.

Together, these results illustrate the complementarity of various transient classes in mapping the PBH parameter space, while highlighting the observational regimes—such as short-duration, cosmologically distant bursts—where meaningful gains in sensitivity can still be achieved. They also underscore the importance of continued improvements in time resolution, detection rate, and multi-wavelength monitoring to extend coverage and reduce uncertainties in the search for PBH dark matter.


\section{Discussion}
\label{sec:discussion}

Our  multi‑wavelength survey of transient lensing opportunities, incorporating revised optical depths, distances, and clumping factors, reveals several important themes for interpreting existing searches and designing future campaigns.

\subsection{Complementarity in Mass Coverage}

The most striking feature of Figure~\ref{fig:mass_abundance} is the clear complementarity between classes in $M_{\mathrm{PBH}}$ coverage:
\begin{itemize}
    \item \textbf{Low‑frequency radio bursts} (FRBs, pulsar giant pulses) probe the lightest PBHs accessible in this framework, with wave–optics limits reaching $M_{\min}\lesssim 10^{-4}\,M_\odot$ for pulsar giant pulses and $\sim 8\times 10^{-3}\,M_\odot$ for FRBs.  
    FRBs dominate abundance sensitivity in this regime thanks to cosmological path lengths and event rates up to $10^5$~yr$^{-1}$ in the SKA‑scale case.
    \item \textbf{High‑energy transients} (GRB spikes, FXTs, TeV flares) cover a broad swath from $M_{\mathrm{PBH}}\sim 4\times 10^{-5}\,M_\odot$ up to $\gtrsim 10^6\,M_\odot$, uniquely probing the high‑mass range beyond the reach of FRBs, albeit with weaker abundance sensitivity.
    \item \textbf{Local low‑frequency bursts} (solar decametric spikes, planetary/UCD cyclotron emission, auroral radiation) overlap in $M_{\mathrm{PBH}}$ with both FRB and high‑energy domains but have negligible cosmological sensitivity due to extremely low optical depths.
\end{itemize}

\subsection{Event Rate vs. Optical Depth Trade‑offs}

As indicated by Eq.~\eqref{eq:fpbh_limit}, $f_{\mathrm{PBH}}$ sensitivity scales with $N_{\mathrm{events}}\tau_{\mathrm{lens}}$.  
FRBs benefit from moderate $\tau_{\mathrm{lens}}\sim 10^{-3}$ combined with unmatched $N_{\mathrm{events}}$, yielding the lowest limits in our sample ($\lesssim 0.3$ for SKA‑class rates).  
Pulsar giant pulses achieve lower $M_{\min}$ but, with $\tau_{\mathrm{lens}}\sim 10^{-8}$ and $\mathcal{O}(10^3)$ events yr$^{-1}$, produce limits two–four orders of magnitude weaker.

High‑energy cosmological sources share $\tau_{\mathrm{lens}}\sim 10^{-3}$ with FRBs but suffer from annual event rates in the tens–hundreds, placing their $f_{\mathrm{PBH}}$ limits in the $10^2$–$10^3$ range.  
Local bursts, even with optimistic clumping factors, have $\tau_{\mathrm{lens}}\lesssim 10^{-20}$, so their limits lie far above unity.

\subsection{ Opportunities and Caveats}

Several developments could improve transient lensing reach:
\begin{itemize}
    \item \textbf{Next‑generation FRB surveys} (SKA, DSA‑2000, CHORD) increasing $N_{\mathrm{events}}$ by 1–2 orders of magnitude would push $f_{\mathrm{PBH}}$ well below $10^{-1}$ in the $8\times 10^{-3}$–$10^2\,M_\odot$ range.
    \item \textbf{Low‑frequency FRB detections} (tens of MHz) would lower $M_{\min}$ by wave–optics scaling, potentially reaching $M_{\mathrm{PBH}}\lesssim 10^{-6}\,M_\odot$.
    \item \textbf{High‑time‑resolution gamma‑ray instruments} could detect $\mu$s‑scale GRB substructure, lowering $M_{\min}$ for high‑energy channels by up to two orders of magnitude.
    \item \textbf{Targeted searches for rare, ultra‑short classes} such as SGR spikes or UCD bursts could fill currently unpopulated regions of the mass–frequency space.
\end{itemize}

{
\paragraph{Caveats (i): clustering, binaries, and extended mass functions.}
Our forecasts have assumed a spatially homogeneous population of PBHs of fixed mass. 
In reality, several effects could alter the lensing phenomenology. 
First, PBHs may exhibit non-trivial clustering, either primordial or induced by halo 
dynamics, which would change the probability of encountering a lens along a given 
sightline. Second, PBHs bound in binaries or small-$N$ associations could introduce 
compound lensing signatures or modify time-delay statistics, particularly in the 
stellar- to intermediate-mass ranges. Finally, an extended PBH mass function would 
smooth the sharp boundaries in $M_{\rm min}$ and $M_{\rm max}$ adopted here, redistributing 
optical depth across a wider range of scales. While a full treatment of these 
effects is beyond the scope of the present work, they represent important caveats 
to keep in mind when interpreting both forecasts and null results from transient 
lensing searches.

\paragraph{Caveats (ii): Role of the clumping factor.}
In our calculations, a clumping factor $\delta \equiv \rho_{\rm local}/\bar\rho_{\rm DM}$ 
was introduced as a simple parametrization of modest PBH overdensities along specific 
sightlines (e.g.\ towards the Galactic plane). This treatment is appropriate when PBHs 
are clustered on scales much larger than their Einstein radius, such that each lens can 
be treated independently. However, if PBHs are so densely packed that the mean separation 
within a clump becomes comparable to the Einstein radius, then collective lensing effects 
arise: the ensemble acts more like a single, extended lens, and the single-PBH optical 
depth approximation breaks down. In this strongly clustered regime, one would need to 
model the lensing by bound PBH clusters or minihalos directly, rather than through an 
effective $\delta$. Our results therefore apply under the assumption of moderate 
clumping ($\delta \sim 5$--10), which captures typical overdensities in Galactic or 
local environments, while more extreme scenarios remain an open direction for future 
work.
}

\subsection{Synergies with Other Probes}

Transient lensing offers a complementary route to PBH constraints:
\begin{itemize}
    \item Microlensing (OGLE, Subaru/HSC) covers overlapping masses but with stellar rather than burst sources.
    \item Gravitational waves constrain the merging fraction of higher‑mass PBHs.
    \item CMB anisotropy and spectral distortion limits probe accretion signatures at high redshift.
\end{itemize}
Its key advantage is model independence: the method depends only on gravitational lensing geometry, not on uncertain astrophysical emission or accretion models.

{
\paragraph{Microlensing of stars and quasars.}
Traditional microlensing searches, in particular those based on long-term monitoring of 
stellar fields and quasars, provide some of the most stringent existing constraints on 
the PBH dark matter fraction over wide mass ranges. For example, the Subaru/HSC survey 
of M31~\cite{1701.02151} rules out $f_{\rm PBH} \gtrsim \mathcal{O}(10^{-2})$ for 
masses around $10^{-11}$--$10^{-6}\,M_\odot$, while OGLE Galactic microlensing analyses 
exclude a dominant PBH population at $\sim 10^{-2}$--$10\,M_\odot$. 
Quasar microlensing, which is sensitive to compact objects near the solar-mass scale, 
similarly indicates that only a subdominant fraction of dark matter can be in the form of PBHs 
in that regime~\cite{2007.10722}. These results occupy and in many cases close large regions 
of the PBH mass–abundance plane, and would appear as stringent exclusion bands overlapping 
with the transient-based reach shown in Figures~1 and~2. 
Our focus in this work has instead been on \emph{transient lensing} of short-duration events, 
where detection relies on temporal or spectral signatures (e.g.\ echoes or interference fringes) 
rather than statistical magnification distributions. The two approaches are complementary: 
microlensing provides the strongest constraints from steady or long-duration sources, while 
transient lensing highlights opportunities in underexplored regimes of parameter space and 
leverages the rapidly growing domain of time-domain astronomy.

\paragraph{Supernovae and gravitational-wave lensing.}
Although not the focus of our survey, it is important to emphasize that 
other classes of non-transient sources also provide powerful PBH constraints. 
Type Ia supernovae, through statistical analyses of Hubble-diagram residuals, 
constrain compact objects heavier than $\sim 0.01\,M_\odot$ to contribute less 
than $\sim 35\%$ of the dark matter at 95\% C.L.~\cite{Zumalacarregui:2017qqd}. 
Strongly lensed SNe Ia may also reveal PBH populations through differential 
microlensing of multiple macro-images. Gravitational-wave transients, on the 
other hand, provide sensitivity to compact lenses in the 
$10 \lesssim M/M_\odot \lesssim 10^5$ range, where wave-optics fringes or 
stochastic diffraction effects can imprint characteristic modulations in 
inspiral signals~\cite{JungShin2019,Basak2021,Zumalacarregui2024LSD}. 
Together, these approaches map onto the same $M_{\rm PBH}$--$f_{\rm PBH}$ 
plane we use here, and complement the electromagnetic transients emphasized 
in this work.

\paragraph{Local solar-system transients.}
We have also included, for completeness, classes of coherent bursts 
arising in the solar system (e.g.\ solar radio spikes, planetary and 
brown dwarf cyclotron bursts). While these sources extend the diversity 
of transient phenomena considered across the electromagnetic spectrum, 
their effective lensing optical depth is extremely small 
($\tau_{\rm lens} \lesssim 10^{-20}$ even for $f_{\rm PBH}=1$). 
As a result, they provide no meaningful constraints on PBH abundance 
unless PBHs are extraordinarily clustered in the local environment. 
These channels should therefore be regarded as illustrative endpoints of 
our methodology, relevant in principle to very low PBH masses or exotic 
clustering scenarios, but not as competitive probes in the mass ranges 
where cosmological and Galactic transients offer sensitivity.

}

\subsection{Roadmap for Future Surveys}

Our results suggest a multi‑pronged strategy:
\begin{enumerate}
    \item \textbf{Maximize event rates} through wide‑field, continuous monitoring across all regimes.
    \item \textbf{Push to lower frequencies} for short‑duration radio transients to reduce $M_{\min}$.
    \item \textbf{Improve time resolution} in high‑energy surveys to access lighter PBHs.
    \item \textbf{Target under‑observed classes} that can populate gaps in Figure~\ref{fig:mass_abundance}.
\end{enumerate}
Pursuing these directions can substantially narrow the remaining PBH dark matter parameter space or, in the most optimistic case, reveal the first transient lensing signature of primordial origin.


\section{Conclusions}
\label{sec:conclusions}

We have carried out the first systematic, multi‑wavelength assessment of astrophysical transient classes with potential for primordial black hole (PBH) lensing searches.  
Our analysis spans over twenty decades in photon frequency and more than fourteen orders of magnitude in PBH mass sensitivity, from low‑frequency radio bursts to TeV $\gamma$‑ray flares.  
By combining wave–optics and time–delay constraints with realistic event rates, distances, clumping factors, and lensing optical depths, we have mapped each transient class into a unified $M_{\mathrm{PBH}}$–$f_{\mathrm{PBH}}$ diagram and summarized their capabilities in two comprehensive tables.

The most stringent projected constraints arise from fast radio bursts (FRBs), which dominate sensitivity in the mass range $\sim 8\times 10^{-3}$–$10^2\,M_\odot$.  
With SKA-scale detection rates, one year of observations could yield $f_{\mathrm{PBH}} \lesssim 0.3$, while current CHIME-scale rates already imply constraints at the level of $f_{\mathrm{PBH}} \sim 3$.
Pulsar giant pulses access slightly lighter PBHs ($M_{\min} \lesssim 10^{-4}\,M_\odot$) but suffer from extremely small optical depths ($\tau_{\mathrm{lens}} \simeq 1.6\times 10^{-8}$), producing one-year limits several orders of magnitude weaker.  
High-energy cosmological transients — including GRB spikes, fast X-ray transients, and TeV flares — expand coverage up to $\gtrsim 10^6\,M_\odot$ but yield $f_{\mathrm{PBH}}$ constraints at the $10^2$–$10^3$ level due to limited event statistics.  
At the opposite extreme, local low-frequency bursts from solar, planetary, or substellar cyclotron emission can probe intermediate PBH masses, but their cosmological optical depths are vanishingly small ($\tau_{\mathrm{lens}} \ll 10^{-20}$), rendering their abundance limits uninformative even under optimistic assumptions for PBH clustering.

These findings define a concrete roadmap for expanding PBH lensing searches:
\begin{enumerate}
    \item Deploy next-generation FRB facilities to push $f_{\mathrm{PBH}}$ constraints below $10^{-1}$ across a wide mass range.
    \item Extend short-duration radio transient monitoring to lower frequencies (tens of MHz) to reduce $M_{\min}$ via wave–optics scaling.
    \item Improve time resolution for high-energy transient observations to shift their sensitivity toward lighter PBHs.
    \item Explore underutilized regions of frequency–duration space by targeting rare or poorly studied transient classes.
    \item Combine transient lensing constraints with complementary PBH probes — microlensing, gravitational waves, and CMB signatures — to achieve comprehensive coverage.
\end{enumerate}

An important cross-cutting consideration is the finite size of transient emission regions, which can suppress lensing signatures in the wave–optics and geometric optics regimes.  
While current constraints are largely unaffected for FRBs and GRBs, source size effects could become critical for future campaigns targeting rare or longer-duration events.  
Careful characterization of source compactness — particularly for radio microstructure and high-energy spikes — will be essential for interpreting null results and for refining the optimal survey design.

With the accelerating growth of time-domain astronomy and multi-wavelength monitoring, the strategy outlined here can be realistically implemented within the coming decade.  
Whether it closes the remaining PBH dark matter windows or reveals the first signatures of primordial lensing, this approach ensures that the viable parameter space for PBH dark matter will be thoroughly and systematically explored.

{

In addition to the transient classes surveyed here, it is worth stressing that 
other probes such as Type~Ia supernova microlensing and gravitational-wave 
lensing provide powerful and complementary constraints on PBH dark matter. 
Supernovae currently set the tightest limits in the $\sim 0.01$--$100\,M_\odot$ 
range, while gravitational-wave diffraction and lensing searches are uniquely 
sensitive to $10 \lesssim M/M_\odot \lesssim 10^5$. Although not analyzed in 
detail in this work, these techniques map directly onto the same 
$M_{\rm PBH}$--$f_{\rm PBH}$ parameter space and should be regarded as 
essential components of a comprehensive PBH search strategy.

Finally, we note that traditional microlensing surveys of stars and quasars 
(such as OGLE, Subaru/HSC, and quasar microlensing analyses) already provide 
some of the strongest existing constraints on the PBH dark matter fraction 
across wide mass ranges. While not the main focus of this work, these results 
map onto the same $M_{\rm PBH}$--$f_{\rm PBH}$ parameter space presented here 
and should be regarded as complementary to the transient-lensing opportunities 
we have surveyed. Taken together, microlensing of steady sources and lensing 
of short-duration transients offer a comprehensive suite of probes that, in 
combination with gravitational waves and other cosmological tests, are rapidly 
closing the viable windows for PBH dark matter.
}


\section*{Acknowledgments}
 This work is partly supported by the U.S.\ Department of Energy grant number de-sc0010107.

\appendix

\section{Optical Depth Calculations for Local, Galactic, and Cosmological Sources}
\label{app:tau_calc}\label{app:optical_depth}

The \emph{lensing optical depth} \(\tau_{\mathrm{lens}}\) is the probability that a given line of sight to a source passes within the Einstein radius of at least one PBH lens.  
For a PBH population making up a fraction \(f_{\mathrm{PBH}}\) of the dark matter in objects of mass \(M\), the comoving number density is
\begin{equation}
    n_{\mathrm{PBH}}(z) = f_{\mathrm{PBH}} \, \frac{\rho_{\mathrm{DM}}(z)}{M} ,
\end{equation}
with \(\rho_{\mathrm{DM}}(z) = \rho_{\mathrm{DM},0} (1+z)^3\) for cosmological sources, or given by a Galactic halo density profile for local sources.

The general optical depth expression is
\begin{equation}
    \tau_{\mathrm{lens}}(M) = \int_{\mathrm{l.o.s.}} dl \ n_{\mathrm{PBH}}(l) \ \pi \left[ y_{\max} \, R_E(l) \right]^2 ,
    \label{eq:tau_general_appendix}
\end{equation}
where \(y_{\max}\) is the maximum detectable impact parameter in units of the Einstein radius and \(R_E(l)\) is the physical Einstein radius at lens position \(l\),
\begin{equation}
    R_E = \sqrt{ \frac{4GM}{c^2} \ \frac{D_L D_{LS}}{D_S} } .
\end{equation}
For circular symmetry, \(\tau_{\mathrm{lens}} \propto y_{\max}^2\) and is nearly mass‑independent for fixed geometry, since \(n_{\mathrm{PBH}} \propto M^{-1}\) while \(R_E^2 \propto M\).

\subsection{Cosmological Sources}
For sources at \(z_S \sim 0.5{-}2\) and a uniform dark matter density, the optical depth is~\cite{Paczynski:1986,Oguri:2018muv}
\begin{equation}
    \tau_{\mathrm{lens}}(f_{\mathrm{PBH}}=1) \simeq \frac{3}{2} \ \Omega_{\mathrm{DM}} \ y_{\max}^2 \ \int_0^{z_S} dz_L \ \frac{(1+z_L)^2}{E(z_L)} \ \frac{D_L D_{LS}}{D_S^2} ,
\end{equation}
where \(E(z) = H(z)/H_0\).  
For \(y_{\max} \sim 1\), \(z_S \sim 1\), and standard \(\Lambda\)CDM parameters, one finds
\[
    \tau_{\mathrm{lens}} \sim (0.5{-}2) \times 10^{-3}
\]
for \(f_{\mathrm{PBH}}=1\).  
Moderate line‑of‑sight overdensities (\(\delta \sim 2{-}3\)) can raise this by a similar factor; strong cluster lenses can push local values to \(\sim 10^{-2}\) but only for rare sightlines.

\subsection{Galactic Sources}
For sources at \(D_S \sim 1{-}10\)~kpc within the Milky Way halo, the optical depth is
\begin{equation}
    \tau_{\mathrm{lens}} = \frac{\pi y_{\max}^2}{M} \ \int_0^{D_S} dl \ \rho_{\mathrm{DM}}[r(l)] \ \frac{4GM}{c^2} \frac{l (D_S - l)}{D_S} ,
\end{equation}
where \(\rho_{\mathrm{DM}}(r)\) is taken from a Galactic halo model (e.g., NFW or Burkert).  
For a local density \(\rho_{\mathrm{DM},\odot} \approx 0.3\ \mathrm{GeV\,cm^{-3}}\) and \(y_{\max} \sim 1\),
\[
    \tau_{\mathrm{lens}}(f_{\mathrm{PBH}}=1) \sim 10^{-6}
\]
for \(D_S \approx 5\ \mathrm{kpc}\).  
Lines of sight through the Galactic center can raise this to \(\sim 10^{-5}\), while PBHs clustered in halo substructure could yield order‑of‑magnitude enhancements for specific directions.

\subsection{Local Sources}
For sources within the Solar System (\(D_S \lesssim 1\ \mathrm{AU}\)),
\begin{equation}
    \tau_{\mathrm{lens}} \approx n_{\mathrm{PBH}} \ \pi (y_{\max} R_E)^2 \ D_S ,
\end{equation}
with \(R_E \propto \sqrt{D_L D_{LS} / D_S}\) becoming extremely small.  
For \(M \sim 1 M_\odot\), \(D_S \sim 1\ \mathrm{AU}\), and \(\rho_{\mathrm{DM}} \approx 0.3\ \mathrm{GeV\,cm^{-3}}\),
\[
    \tau_{\mathrm{lens}}(f_{\mathrm{PBH}}=1) \lesssim 10^{-12}.
\]
Even extreme local overdensities (e.g., PBHs bound to the Solar System) would be required to raise this to detectable levels; enhancements by factors \(\delta \gtrsim 10^6\) are implausible unless PBHs form bound microhalos in the immediate vicinity.

\subsection{Summary of Typical Ranges}
In the absence of extreme overdensities, reasonable ranges for \(\tau_{\mathrm{lens}}(f_{\mathrm{PBH}}=1)\) are:
\begin{equation}
\begin{aligned}
    &\text{Cosmological:} & 5\times 10^{-4} &\lesssim \tau_{\mathrm{lens}} \lesssim 3\times 10^{-3}, \\
    &\text{Galactic:} & 3\times 10^{-7} &\lesssim \tau_{\mathrm{lens}} \lesssim 3\times 10^{-5}, \\
    &\text{Local:} & 10^{-14} &\lesssim \tau_{\mathrm{lens}} \lesssim 10^{-12}.
\end{aligned}
\end{equation}
The upper ends correspond to favorable geometries or moderate clumping; substantially higher values would require unobserved extreme PBH overdensities.

 \subsection*{Declaration of generative AI and AI-assisted technologies in the writing process.}
 During the preparation of this work, the author used the {\tt undermind.ai} tool 
to assist in surveying the relevant literature and compiling a list of references. 
All retrieved material was subsequently reviewed and edited by the author, who 
takes full responsibility for the accuracy and integrity of the final published article.

\bibliographystyle{JHEP}

\bibliography{references}

\end{document}